# Farthest Point Sampling in Property Designated Chemical Feature Space as a General Strategy for Enhancing the Machine Learning Model Performance for Small Scale Chemical Dataset

Yuze Liu[a, b], Xi Yu[a, b*]

[a] Key Laboratory of Organic Integrated Circuit, Ministry of Education & Tianjin Key Laboratory of Molecular Optoelectronic Sciences, Department of Chemistry, School of Science, Tianjin University, Tianjin 300072, China

[b] Collaborative Innovation Center of Chemical Science and Engineering (Tianjin), Tianjin 300072, China

[*] Email: xi.yu@tju.edu.cn

**ABSTRACT:** Machine learning model development in chemistry and materials science often grapples with the challenge of small-scale, unbalanced labelled datasets, a common limitation in scientific experiments. These dataset imbalances can precipitate overfitting and diminish model generalization. Our study explores the efficacy of the farthest point sampling (FPS) strategy within targeted chemical feature spaces, demonstrating its capacity to generate well-distributed training datasets and consequently enhance model performance. We rigorously evaluated this strategy across various machine learning models, including artificial neural networks (ANN), support vector machines (SVM), and random forests (RF), using datasets encapsulating physicochemical properties like standard boiling points and enthalpy of vaporization. Our findings reveal that FPS-based models consistently surpass those trained via random sampling, exhibiting superior predictive accuracy and robustness, alongside a marked reduction in overfitting. This improvement is particularly pronounced in smaller training datasets, attributable to increased diversity within the training data's chemical feature space. Consequently, FPS emerges as a universally effective and adaptable approach in approaching high performance machine learning models by small and biased experimental datasets prevalent in chemistry and materials science.

## 1. INTRODUCTION

Machine learning (ML) has significantly advanced the fields of chemistry and material science[1-4], propelling the study of cheminformatics and enabling rapid structure-property prediction and design[5-11]. However, the inherent requirement on the extensive dataset for ML study raised significant challenge for practical application of ML in experimental science[12]. Often, labelled experimental chemical and material datasets are limited in size and coverage, and most significantly imbalanced[13, 14], due to constraints in data acquisition, like time, cost, and technical barriers. Consequently, ML models trained on these datasets, which are frequently subsampled randomly for training and testing, are prone to overfitting and exhibit diminished generalization capabilities due to the imbalanced nature of the data, where certain types of observations are disproportionately represented compared to others. The complexity of these challenges is further magnified by the high dimensionality of chemical data and the intricate nature of chemical scenarios.

To mitigate these issues, various sampling methods have been employed to achieve data balance and curtail the risk of overfitting[15, 16]. Conventional methods like oversampling and under-sampling[17, 18] directly manipulate the dataset's size to address class imbalances but can lead to loss of information or overfitting, while stratified sampling[19] maintains the proportion of classes but doesn't necessarily enhance data diversity. Advanced methods bring nuanced solutions with their own trade-offs. The Genetic Algorithm (GA) method[20], while optimizing data point selection for diversity, can be computationally intensive and may require careful tuning to avoid converging on suboptimal solutions. Active learning[21-24] effectively refines models iteratively by selecting informative data, yet this comes with the cost of increased computational demands due to continuous model updates. Another promising avenue is leveraging non-labeled pretrained models, such as transformers[25-27], followed by post-fine-tuning using labeled chemical data. However, while these pretrained transformer models can capture extensive knowledge from vast amounts of unlabeled data, they come with substantial computational and resource demands. This makes them challenging to deploy in dynamic chemical scenarios where rapid adaptability and cost-efficiency are paramount.

In this paper, we introduce a general strategy to address the challenge, the farthest point sampling (FPS) in property designated chemical feature space. FPS is a sampling method tailored for high-dimensional spaces. It operates on the assumption that the distribution of structures in diversity spaces is important, which provides the prior information to guide its sampling. The method selects samples in the feature space that are furthest apart, effectively capturing the essence of the entire dataset with a minimal number of samples. Although FPS has been utilized in progressive image sampling[28], point cloud networks[29], and feature selection[30], its application chemical dataset for machine learning has not been explored. In this study, we employed FPS in the property designated chemical feature space (FPS-PDCFS) to partition databases of physicochemical properties such as standard boiling points and enthalpy of vaporization. In our study, ML models, including artificial neural networks, support vector machines, and random forests, utilizing FPS consistently outperformed those based on Random Sampling (RS), showcasing enhanced predictive accuracy, stability and markedly reduced overfitting, and a robust resilience to limited sample sizes. Our findings underscore that within a property-designated high-dimensional feature space, FPS adeptly identifies unique characteristics and preserves the training set's diversity. Such an approach substantially elevates the efficacy of ML models in predicting molecular properties by ensuring a holistic and balanced portrayal of the chemical feature landscape. Consequently, FPS-PDCFS positions itself as a versatile sampling tool, enhancing both the quality of small, skewed chemical datasets and the predictive capability of chemical ML models.

## 2. METHODS

### 2.1 Physicochemical Database and Machine Learning Model

The thermodynamic and physical properties dataset we used for the study is sourced from the Yaws' handbook[31] and from internet databases such as PubChem[32]. The dataset encompasses structurally diverse compounds, including hydrocarbons and heterocycles, and features properties like boiling point, enthalpy of vaporization, and critical properties. These properties are utilized to compare the performance of FPS and random sampling. Additional details regarding the datasets can be found in the Supporting Information (SI), section 1.

We have selected several well-known machine learning models, such as Artificial Neural Networks (ANN), Support Vector Machines (SVM), Kernel Ridge Regression (KRR), and k-Nearest Neighbours (KNN). These methods are frequently employed for regression and classification tasks aimed at predicting material or molecular properties. The training super parameters of these models are optimized through grid search, and all comparisons are conducted using the same training parameters, which are detailed in the SI. Molecular features (descriptors) were calculated using AlvaDesc[33]. A set of interpretable descriptors comprising a series of constituent descriptors and topological indices were chosen as outlined in our previous work[34]. These interpretable descriptors consistently demonstrate stability and high accuracy in modelling physicochemical properties of organic chemicals.

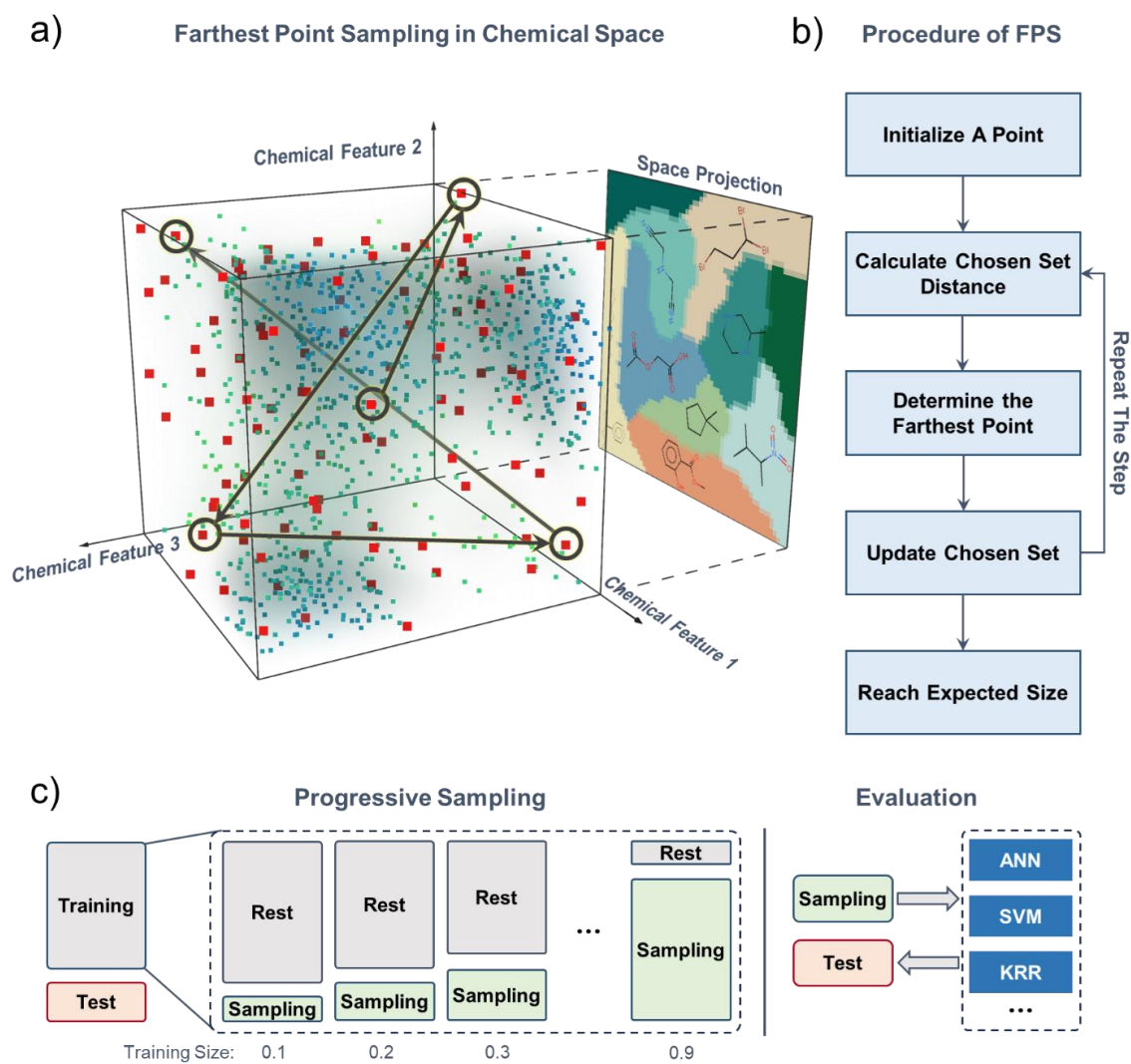

**Figure 1.** (a) A graphic illustration of the farthest point sampling in chemical space. (b) A Flow chart of FPS Procedure. (c) Sampling process used in this study. The initial dataset was partitioned into training and test sets randomly. The training set was then further bifurcated into a "sampling set" and a "rest set" of different ratio by sampling using different strategies. Various machine learning models were next trained on the sampling set and subsequently validated on the test set.

2.2 Sampling Method and Benchmark

Random sampling is a straightforward process involving the selection of a random subset from the dataset. On the other hand, FPS necessitates a more structured approach, requiring execution within a pre-defined chemical feature space. We implemented FPS within the space defined by molecular descriptors utilized in our machine learning model. This choice was informed by our discovery that FPS could augment model performance when executed within a feature space correlated with the target properties, a phenomenon we will elucidate in subsequent sections. In **Fig. 1a**, FPS is demonstrated to sample within a chemical feature space articulated by molecular descriptors. Diverse points within this space represent various categories of molecules from the chemical database, each occupying distinct positions and exhibiting varying densities. FPS is adept at uniformly sampling across this space, ensuring a representative collection of molecules from each category. As depicted in **Fig. 1b**, the process of FPS sampling encompasses the following steps:

1) randomly select an initial point from the dataset.

2) compute the distances from all other points to this initial point and select the farthest point as the second sampled point.

3) for each new sampled point, compute the minimum distances from all unsampled points P to all previously sampled points S, which is

$$d_{min}(P,S) = min_{s \in S} \sqrt{(P_x - s_x)^2 + (P_y - s_y)^2 + (P_z - s_z)^2}$$

where $(P_x, P_y, P_z)$ are the coordinates of point $P$ and $(s_x, s_y, s_z)$ are the coordinates of point $s$. The point with the maximum distance of $d_{min}(P,S)$ is selected as the next sampled point.

4) Step 3 is reiterated until the chosen set attains the desired size. Consequently, the FPS sampling set comprises points that are maximally distant from one another.

To comprehensively compare model performance across different sampling methods, we employ subsampling within the dataset as shown in **Fig. 1c**. The initial dataset was partitioned into training and test sets randomly. The training set was then further bifurcated into a "sampling set" and a "rest set" of different ratio by using different sampling methods. Various machine learning models were next trained on the sampling set and subsequently validated on the test set, enabling us to assess the efficacy of the sampling strategies. We employed 5-fold cross-validation (CV) across multiple subsets of the data for enhanced reliability of our assessment of model accuracy and generalization capabilities. We defined $MSE_{Train}$ as the mean squared error of the average training set throughout the 5-fold process. The training sizes were incrementally adjusted from 0.1 to 1 of the total dataset, with an increasement of 0.1. Note here only the sampled portion of the training data were used in the training process until the training size reached 1. We will show below that models by FPS-PDCFS at small size of the training data performs better than full training

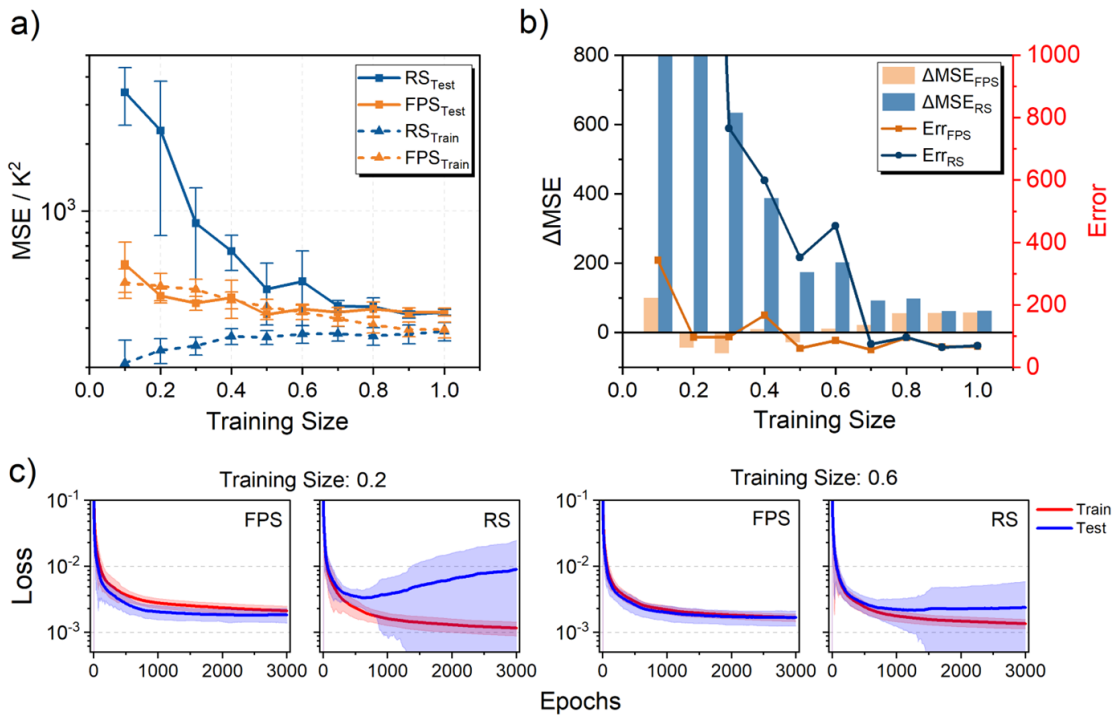

**Figure 2.** (a) MSE of training and test sets of the ANN model under the Farthest Point Sampling method (yellow) and Random Sampling (blue) at different training sizes on the boiling point dataset. (b) The difference between Test MSE and Train MSE (ΔMSE), along with the errors associated with FPS and RS at various training sizes was analyzed. A training size of 0.6, which showed the lowest ΔMSE, indicates the least overfitting. (c) Loss curve for training and test set during the training procedure of ANN, by FPS and RS at training size of 0.2 and 0.6, the large shading area at 0.6 indicates higher overfitting risk and variance in predictive performance with random sampling.

dataset, which is a strong indication of the imbalance of the chemistry dataset.

## 3. RESULTS AND DISCUSSION

Following the procedure shown in **Fig. 1**, we began a systematic evaluation of the FPS strategy's effectiveness, comparing it directly with the effectiveness of standard random sampling. We chose to use an ANN as our benchmark because it is one of the most widely used and well-established machine learning methods, making it a solid foundation for assessing the effectiveness of the FPS strategy.

**Fig. 2** presents a comparative analysis of the ANN model's performance, contrasting the use of FPS and RS across various training sizes within the boiling point dataset. A detailed inspection of **Fig. 2a** and **Fig. 2b** reveals a pronounced disparity in MSE between the training and test sets under RS, particularly at smaller training sizes. This disparity signals significant overfitting, a common issue in models trained on small and imbalanced datasets.[35] Yet, as the training size increases, this tendency towards overfitting gradually diminishes. In contrast, the FPS-enhanced ANN model demonstrates a remarkable alignment of MSEs for both training and test sets across a majority of training sizes, as depicted in **Fig. 2b**. This alignment indicates a consistent reduction in overfitting, even with smaller training sets. Notably, at a training size of 0.6, both the test set MSE and the ΔMSE reach their lowest points, suggesting that the FPS-enhanced ANN model not only achieves comparable accuracy to the RS-based model using the full dataset, but also shows reduced overfitting and enhanced generalizability. As the training size grows, the performance of FPS progressively converges with that of RS, which is expected as the characteristics of the training sets for both FPS and RS become increasingly similar.

Furthermore, the negative ΔMSE values at training sizes of 0.2, 0.3, and 0.5 indicating superior test set performance compared to the training set, emphasizing the FPS strategy's role in bolstering model generalizability and predictive reliability.

Remarkably, the MSEs for both training and test sets remain closely matched even at a minimal training size of 0.1 and 0.2, with the test set showing significantly lower MSE, indicative of FPS's efficacy in achieving superior performance. This observation supports the notion that FPS facilitates the attainment of high-quality model outcomes with minimal data, significantly lowering the data requirements and associated costs in machine learning. Given the challenges and high expenses of chemical data collection, this efficiency is particularly valuable for applying machine learning in chemistry and materials science, thereby paving the way for more accurate and cost-effective research methodologies.

The enhanced stability, robustness and reduced overfitting attributed to the FPS strategy are further supported by the learning curve shown in **Fig. 2c (See Supporting Information, Section 4, Figure S3, for comprehensive learning curves)**. It becomes apparent that FPS leads to greater stability and reduced overfitting in the learning process compared to random sampling, particularly at a training size smaller than 0.3.

In our continued exploration, we expanded the scope of our investigation to assess the effectiveness of FPS across a variety of machine learning models, including ANN, SVM, KRR, KNN, Random Forest (RF) and Extreme Gradient Boosting (XGB), **(as detailed in Supporting Information Figures S4 and S5)**.

Building on the earlier discussion, the successful application of FPS necessitates an appropriate feature space. To delve deeper into this aspect, we conducted a thorough evaluation of FPS's performance across a range of feature spaces. Specifically, we scrutinized three different feature space configurations, each characterized by a unique set of features:

**Set A - Interpretable Descriptors**, based on the physical principles linking molecular features to their boiling points, which are also the descriptors used in our ANN model, aiming to reflect the fundamental physical characteristics of the molecules.

**Set B - Regression-Derived Descriptors**, derived from regression analysis, specifically focusing on the correlations between descriptors and the target property, which may not be immediately apparent but are statistically significant within the dataset.

**Set C - Randomly-Selected Descriptors**, a random selection of descriptors from a comprehensive pool of 1600 descriptors available in alvaDesc, offering a control set with respect to Sets A and B.

**Table 1. The Descriptor Type of chosen descriptors for FPS feature space**

| Descriptor Type | Set A[a] | Set B[b] |
|---|---|---|
| Molecular Properties | √ | √ |
| Functional Group | √ | |
| Constitutional Indices | √ | √ |
| Topological Indices | √ | √ |
| Structural Parameter | √ | |
| Information Indices | | √ |
| 2D Autocorrelations | | √ |
| Burden Eigenvalues | | √ |
| P_VSA-like Descriptors | | √ |
| Pharmacophore Descriptors | | √ |
| Charge Descriptors | √ | √ |
| Drug-like Indices | | √ |
| MDE Descriptors | | √ |
| Walk and Path Counts | | √ |

[a] Set A: Interpretable Descriptors Set

[b] Set B: Regression-Derived Descriptors

The types of descriptors utilized in each configuration are summarized in **Table 1**. For a more detailed exploration of all the descriptors included in this study, refer to the **supporting information provided in Tables S1, S2, and S3**.

In evaluating the effect of the feature spaces, different feature sets were employed only during the FPS, while the ANN model was kept the same.

We can see from **Fig. 3** that FPS with Set A achieved the best performance, indicated by the lowest MSE across all tested feature spaces. For Set B, though manifested a substantial value at a 0.1 training size, MSE of the model at Set B decreased markedly as the training size increased, matching the performance by Set A beyond 0.2. Similar to the scenario in

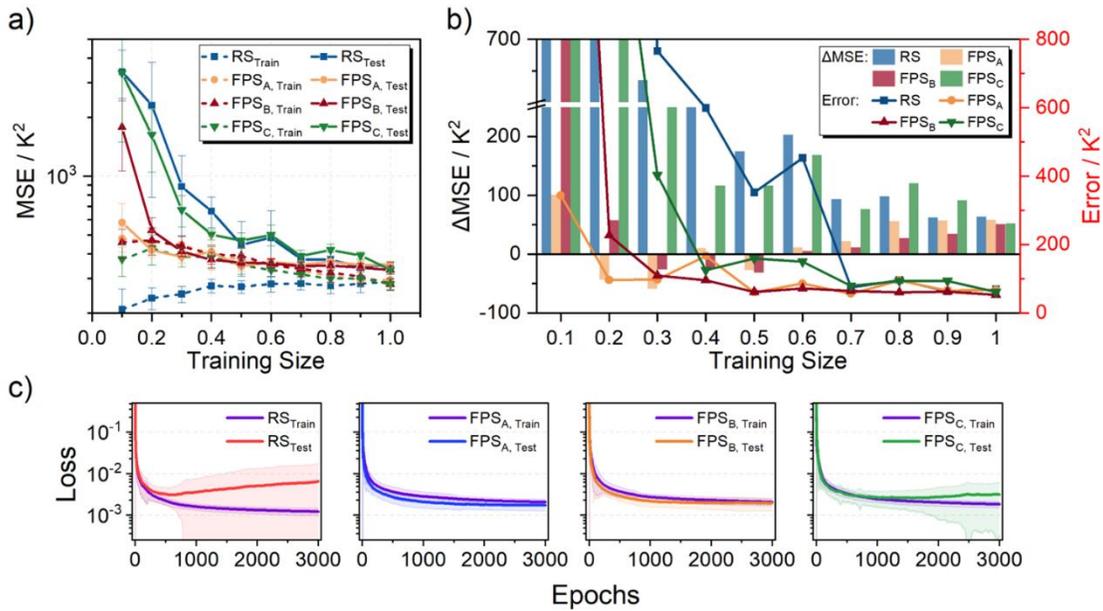

**Figure 3.** (a) MSE of training and test sets by random sampling (red), FPS in the interpretable space (blue), regression space (yellow), and casually selected space (green). (b) The ΔMSE along with the errors under different sampling methods at various training sizes was analysed. (c) Training and Test loss curves with different sampling methods during the training process at a training size of 0.3. The test loss curves for FPS with Set A and B remain more stable and do not show a significant increase as epochs grows. This behaviour suggests that sampling in these feature spaces contributes to model stability during training.

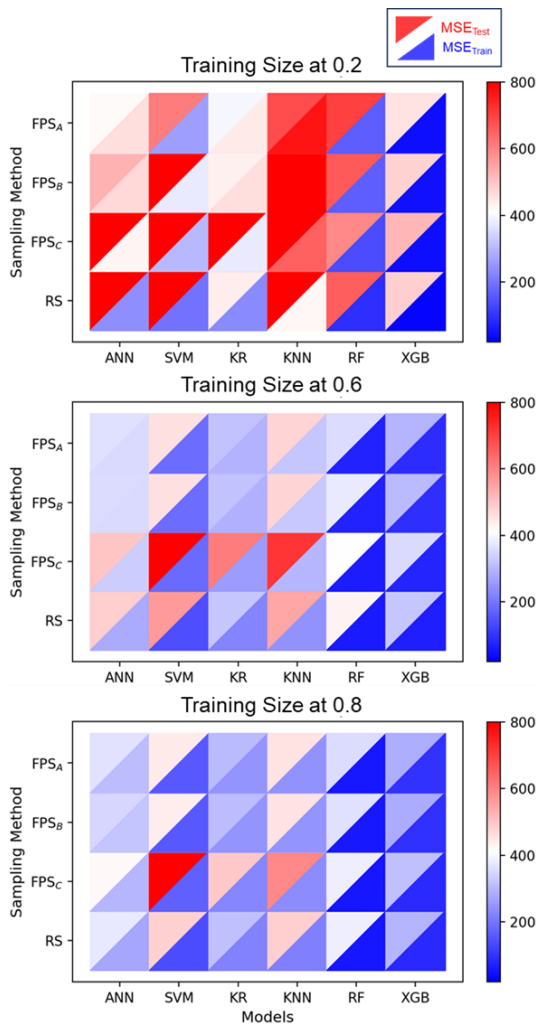

**Figure 4.** The heatmap illustrates the MSE of training and test set under various sampling methods by different machine learning models at training sizes of 0.2, 0.6 and 0.8, with the upper triangle legend representing test set MSE, and the lower triangle legend representing training set MSE. The MSE of the training and test sets for FPS with Set A and B display more similar color patterns across various machine learning models, indicates less overfitting is achieved in these models by implementing the FPS strategy in the designated feature spaces.

Set A, Set B also displayed minimal overfitting at smaller training sizes, with optimal performance noted at a training size of 0.6-0.7, suggesting a best balances at the training dataset at this size, As the training size increases beyond the 0.6-0.7 range, the trend in Set B mirrors that in Set A, where FPS gradually starts to exhibit characteristics of RS, leading to increased overfitting. In contrast, both RS and FPS by Set C underperformed among all the sampling strategies, with FPS by Set C performed similarly with RS. This underscores the critical role of feature space where FPS was performed. Learning curves of different sampling strategies at a training size of 0.3, shown in **Fig. 3c**, demonstrated an early onset of overfitting with the loss of training and testing diverge after tens of training epochs. On the other hand, training under FPS with Set A and B underwent a co-decline of loss of both training and test set and maintained a stable low overfitting in all the training steps. While training by FPS with Set C showed a clear overfitting as well at training step over 2000. These findings underscore FPS's ability to consistently enhance model performance, particularly effective in feature spaces well-correlated with the target properties. Different machine learning models utilizing FPS across various feature spaces were also benchmarked **(see Supporting Information Figure S6)**.

**Fig. 4** illustrates the MSE across different sampling methods and machine learning models, as represented by three heatmaps corresponding to specific training size: 0.2, 0.6, and

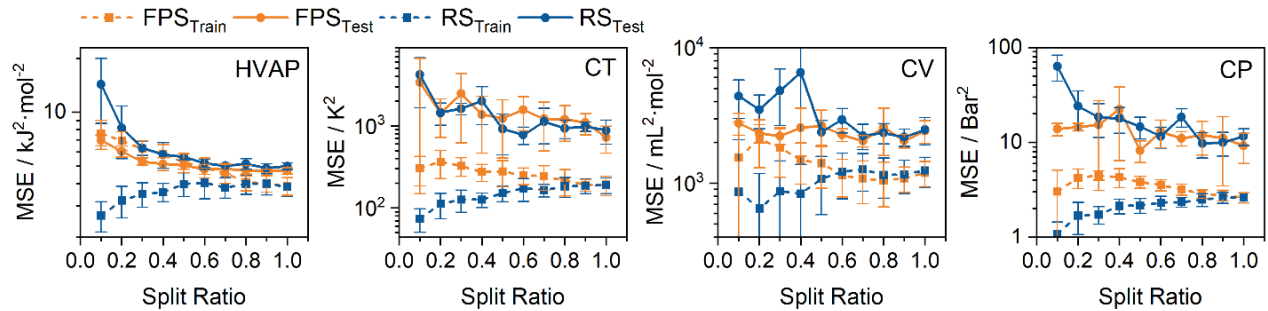

**Figure 5.** Comparison of MSE for training and test sets in the ANN model using FPS (yellow) and RS (blue) across various training sizes on physicochemical datasets. These datasets include the enthalpy of vaporization (HVAP), critical temperature (CT), critical volume (CV), and critical pressure (CP).

0.8. In these heatmaps, the upper and lower triangles denote the test and training sets, respectively. A bluer color indicating lower MSE, while a less pronounced color contrast between the two sets suggests lower overfitting. At a training size of 0.2, not all models exhibit low MSE on the test set, but notably, within the ANN models, those using FPS-A show a relatively low MSE on the test set with a small deviation from the training set MSE when using RS, indicating reduced overfitting. As the training size increases to 0.6, the trend of diminished overfitting becomes more apparent across most models. By the time the training size reaches 0.8, the differences in MSE between FPS and RS methods diminish, with the MSE values of the models closely aligning. It is discernible that ANN, SVM, KR, and KNN all benefit from FPS, as they exhibit relatively low MSE and a minor performance difference between the training and test sets. However, RF and XGB exhibit a distinct behavior where the heatmap reveals no significant distinction between different sampling methods and feature spaces for RF and XGB models, which can be attributed to their nature as ensemble models. These models inherently employ bootstrap sampling, creating multiple data subsets, each embodying a level of random variation, to train various models within the ensemble. The intrinsic bootstrap sampling technique utilized by RF[36] and XGB[37] promotes a measure of diversity and robustness in training process, potentially diminishing the relative benefit derived from the structured approach of FPS[38]. Nonetheless, models implementing the FPS strategy still exhibit advantages over RF and XGB with bootstrap sampling, particularly in terms of reduced overfitting and improved stability.

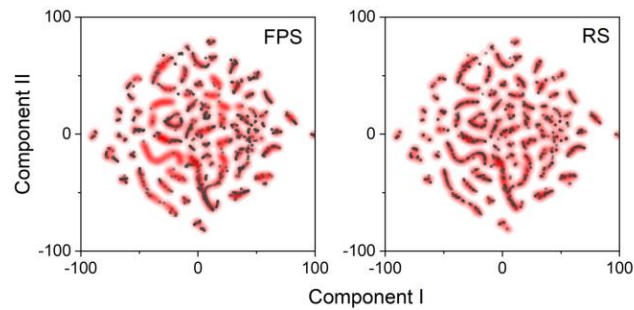

**Figure 6.** -SNE Plot depicting the distribution of boiling point data points. Probability density illustrates the original data distribution, while grey points represent samples obtained through FPS and RS methods. Points sampled by FPS tend to occur more frequently in areas that are difficult to cluster, with fewer appearing in zones of well-defined clustering.

Moreover, we have expanded our study to encompass a variety of physicochemical property databases, including the Enthalpy of Vaporization (HVAP), as well as critical properties such as critical temperature and critical volume, to further assess the effectiveness of FPS. The performance of the ANN model under FPS and RS across these four types of physicochemical property databases is elaborated in **Fig. 5**. Notably, when the training size is small, FPS achieves a lower test MSE and overfitting compared to RS in all HVAP, critical volume (CV), and critical pressure (CP), indicating its greater resistance to overfitting with limited training data. However, it's important to recognize that in HVAP, critical temperature (CT), CV, and CP, the chemical spaces weren't as finely adjusted as in the case of boiling point, which could explain the slightly lesser improvement of FPS over RS in these datasets compared to the boiling point scenario. Additionally, we have evaluated the performance of other machine learning models on these physicochemical property databases, **(as detailed in Supporting Information Figures S7, S8, S9 and S10)**. These models uniformly demonstrate that FPS offers enhanced robustness, predictive accuracy, and consistent reduction in overfitting.

To illustrate how FPS achieves balanced sampling in the data space, encapsulating pivotal chemical structures requisite for nuanced model learning, we employed a 2-dimensional t-SNE plot[39] (t-distributed Stochastic Neighbor Embedding) for Set A shown in **Fig. 6** In this visualization, each point represents a specific molecule from the dataset, and the spatial relationships between points on the plot approximate the relationships between those molecules in the original high-dimensional feature space. The distinct clustering of molecules within this plot underscores the granularity of structural differentiation attained through FPS, showcasing its ability to select a diverse array of structures, including those not well-clustered. The structural diversity by FPS fortifies model learning accuracy - an attribute conspicuously absent in RS, as evidenced by the sampling density plot, and affirmed its superiority over RS in handling complex, high-dimensional datasets.

## 4. CONCLUSION

In this study, we applied the Farthest Point Sampling (FPS) method within a designated chemical feature space to tackle the challenges tied to small, imbalanced datasets prevalent in chemistry and material science domains. Extensive testing across various machine learning models and physicochemical datasets demonstrated that FPS significantly enhances the diversity of training data. This leads to improved predictive accuracy and reduced overfitting in nearly all tested models. A critical discovery is that the success of FPS is tethered to its application within a suitable chemical feature space resonating with the target property, and it works particularly well when applied to ANN with proper pre-knowledge descriptors tethered to the target properties. Consequently, FPS emerges as a versatile, pragmatic strategy, poised to enhance both the quality of small, skewed chemical datasets and the predictive capability of chemical machine learning models, paving the way for more accurate and reliable structure-property predictions in chemistry and material science study with minima data set and reduced cost.

## ASSOCIATED CONTENT

**Supporting Information** includes details of the database, descriptors and model parameters, as well as additional data analyses and is available on http://pubs.acs.org.

## AUTHOR INFORMATION

### Corresponding Author

*Email: xi.yu@tju.edu.cn

### Notes

The authors declare no competing financial interest.

## ACKNOWLEDGMENT

This work was supported by the National Natural Science Foundation of China (21973069, 21773169), 2021 Subsidized Project of Tianjin University Graduate Education Special Fund (B2-2021-003), Shanghai Boronmatrix Advanced Materials Technology Co. Ltd. and the Fundamental Research Funds for the Central Universities.

Farthest Point Sampling was found significantly enhanced training data diversity in small, unbalanced datasets in chemistry and materials science, therefore consistently improved predictive accuracy, and model robustness while reducing overfitting. FPS proves particularly effective for small datasets, greatly reducing data acquisition costs and improving machine learning model performance in chemical and material studies.

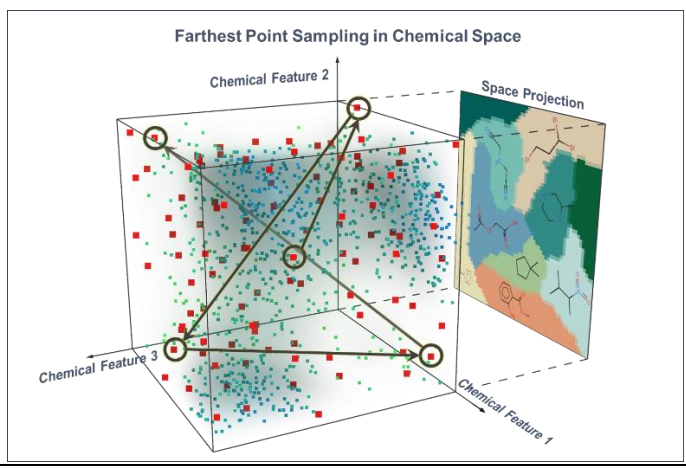